\documentclass[aps,pre,twocolumn,floats,showpacs,amsmath,amssymb,floatfix]{revtex4}
\usepackage{graphicx}
\usepackage{color}

\begin{document}              

\title{Emerging attractors and the transition from dissipative to
conservative dynamics}

\author{Christian
S. Rodrigues\footnote{Electronic Address: c.rodrigues@abdn.ac.uk}}
\affiliation{Department of Physics, King's College, University
of Aberdeen - Aberdeen AB24 3UE, UK.}
\affiliation{Institute for Complex Systems and Mathematical Biology,
King's College, University of Aberdeen - Aberdeen AB24 3UE, UK.}

\author{Alessandro P. S. de Moura} \affiliation{Department of Physics,
King's College, University of Aberdeen - Aberdeen AB24 3UE, UK.}
\affiliation{Institute for Complex Systems and Mathematical Biology,
King's College, University of Aberdeen - Aberdeen AB24 3UE, UK.}

\author{Celso Grebogi} \affiliation{Department of Physics, King's
College, University of Aberdeen - Aberdeen AB24 3UE, UK.}
\affiliation{Institute for Complex Systems and Mathematical Biology,
King's College, University of Aberdeen - Aberdeen AB24 3UE, UK.}

\date{\today}

\begin{abstract}


The topological structure of basin boundaries plays a fundamental role
in the sensitivity to the final state in chaotic dynamical
systems. Herewith we present a study on the dynamics of dissipative
systems close to the Hamiltonian limit, emphasising the increasing
number of periodic attractors, and on the structural changes in their
basin boundaries as the dissipation approaches zero. We show
numerically that a power law with nontrivial exponent describes the
growth of the total number of periodic attractors as the damping is
decreased. We also establish that for small scales the dynamics is
governed by \emph{effective} dynamical invariants, whose measure
depends not only on the region of the phase space, but also on the
scale under consideration.  Therefore, our results show that the concept of
effective invariants is also relevant for dissipative systems.

\end{abstract}


\maketitle


\section{\label{sec.I} Introduction}

\indent Many dynamical processes have been shown to possess coexisting
metastable states, and their dynamics can be highly sensitive to the
initial conditions. Some examples include neural
behaviour~\cite{neural, neural2}, rain events~\cite{kimPRE02},
earthquakes' dynamics~\cite{kimPRL02, chaos04}, effective fractal dimension of energy levels~\cite{wang97}, among others. Their correct interpretation
requires an understanding of dynamical systems with very low
dissipation, lying in between the strongly dissipative limit and the
conservative one. In spite of the importance of this problem,
there are few systematic studies of the low-dissipation limit and the
transition from dissipative to conservative dynamics.  This is the
problem we address in this paper.

In Hamiltonian systems, chaotic regions typically coexist with regions
of regular motion around the marginally stable periodic orbits, also known as
Kolmogorov-Arnold-Moser (KAM) invariant tori. Chaotic trajectories
have intermittent behaviour and spend long times sporadically near the
border of regular islands. This stickiness to the KAM tori makes their
dynamics fundamentally different from that of hyperbolic chaotic
systems, since even small islands can exert great influence on the
global dynamics~\cite{motter05, meiss83, altmann06}. On the other hand,
strongly dissipative systems are characterised as having only one or
few attractors to which all the initial conditions eventually
converge.  These two regimes could not be more different from each
other; however, if the dissipation decreases to zero, the phase space
structure of the dissipative systems must evolve in such a way as to
approach the complex hierarchical organisation found in Hamiltonian
systems. How this transformation occurs is far from obvious.

Despite the existence of extensive investigations on strongly
dissipative and on Hamiltonian systems, very little is known about
properties of systems close to the border between dissipative and
conservative dynamics, where many coexisting periodic attractors are
present~\cite{feudel96}. In this paper we investigate the growth of
the number of periodic attractors and how the topology of the phase
space evolves when the dissipation is reduced. We obtain numerically a
power law describing the growth of the number of periodic attractors
as the damping approaches zero. When the system is close to the
Hamiltonian limit, we find that high-period periodic attractors become
increasingly important in the system's dynamics, in contrast to the
case with high dissipation. These \textit{emerging attractors}
proliferate ever more rapidly as the dissipation decreases. Moreover,
we argue that although this number can be extremely high for small
dissipation, we expect it to be finite (for non-zero dissipation), as
conjectured recently~\cite{palis1, palis2}. The power law of the
number of attractors with dissipation, found in this work,
corroborates this conjecture. Another important finding in this paper
is that for low levels of damping the dynamics is characterised by
\textit{effective invariants}, a concept previously used in the context
of non-hyperbolic dynamics of Hamiltonian systems~\cite{motter05}. We
study specifically the \textit{effective fractal dimension}, which we
show to depend on the scale and also on the region of the phase
space. Therefore we show that effective dynamics is relevant also for
dissipative systems, and not only for Hamiltonian ones.

This paper is organised as follows. We start, with Sec.~\ref{sec.N-of-Att}, 
where we introduce the dynamical
systems used in our analysis and revisit the
problem of the number of coexistent periodic attractors. It is
followed, Sec.~\ref{sec.Topology}, by a discussion regarding the
topology of the phase space in the low-dissipation limit. In
Sec.~\ref{sec.Effective} we introduce effective invariants and define
the effective fractal dimension for weakly dissipative
systems. Finally, Sec.~\ref{Sec.conclu} brings our conclusions.
In the appendix, we bring up some formal definitions. In
particular, we recall a definition of \textit{periodic
  attractors}, and discuss its implications for both 
experimental and numerical realistic investigations, where one
typically is only able to make use of limited precision. We also
recall a definition of \textit{basin of attraction}, which
is coherent  with the definition of periodic attractors
used here.



\section{Number of Periodic Attractors}\label{sec.N-of-Att}

\subsection{Single Rotor Map}

A paradigmatic system that allows the investigation of the transition
from the dissipative to the conservative limit is the \textit{single
rotor map}, obtained from the mechanical pendulum kicked periodically
at times $nT$, $n \in \mathbb{N}^{+}$, with force
$f_{0}$~\cite{Zaslavskii1978}:
\begin{equation}
\label{eq.single-rotor-map}
\begin{split}
x_{n+1} &= x_{n} + y_{n}(mod 2\pi )\\
y_{n+1} &= (1-\nu)y_{n} + f_{0}\sin(x_{n} + y_{n}),
\end{split}
\end{equation}
where $x \in [0,2\pi]$ corresponds to the phase, $y \in \mathbb{R}$ to
the angular velocity, and $\nu \in [0,1]$ is the damping
parameter. For $\nu = 0$, the well known area preserving standard map
analysed by Chirikov~\cite{Chirikov} is recovered. For $\nu = 1$, we
obtain the circle map with zero rotation number.

Previous work has considered the question of the number of
attractors for this system~\cite{feudel96}. They have derived an
analytical expression for the number of period-1 primary attractors, that have not undergone period doubling bifurcation:
\begin{equation}
\label{eq.urike-attrac}
N_{p1} = 2 I\left( \frac{f_{0}}{2\pi \nu}\right) + 1,
\end{equation}
where $I(\cdot)$ denotes the integer part of the expression in
brackets. As
pointed out in Ref.~\cite{feudel96}, numerical detection of attractors
with high period is a difficult task. The reason is the extremely
small size of their basins of attraction, what requires computation
using very fine grids. Adding to this, they have short lifetime in the
space of parameters. Therefore it is necessary to follow a great
number of initial conditions in order to be able to detect them.
\begin{figure}[tb]
\includegraphics[width=.95\columnwidth]{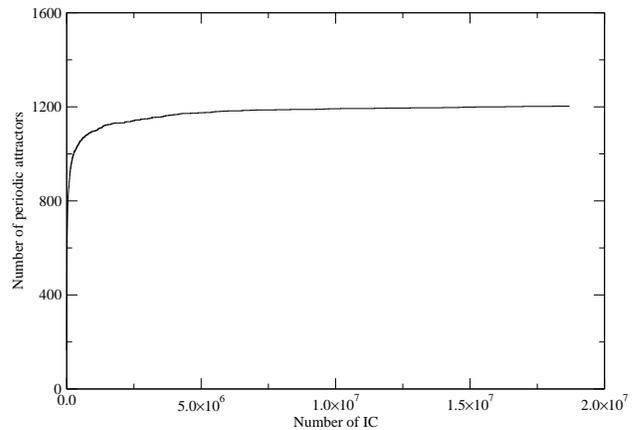}
\caption{Total number of periodic attractors detected for a specific
value of the dissipation $\nu$, as a function of the number of initial
conditions randomly chosen in $\boldsymbol{\Lambda}$ and evolved
according to the map. The parameters for the map
(\ref{eq.single-rotor-map}) were $f_{0} = 4.0$, and $\nu = 0.0035$.}
\label{fig.condxattrac}
\end{figure}

Here we overcome this problem. Instead of choosing an \textit{a
  priori} fixed number of initial conditions in a preset grid, as
previously done in Ref.~\cite{feudel96} and elsewhere, we fix the
non linearity $f_{0}=4.0$ for a given damping $\nu$ and then iterate an
ensemble of randomly chosen initial conditions with uniform
distribution on a bounded region of the phase space. Therefore, the
chance for an attractor to be found does not depend on our particular
choice of the grid.  For the dissipative case the dynamics takes
place in the cylinder $\boldsymbol{\Gamma} = [0, 2\pi] \times
\mathbb{R}$ and the area where the initial conditions are chosen
depends on the damping. Since $\nu>0$, from the second equation in the
map (\ref{eq.single-rotor-map}), one gets $|y_{k+1}| \leq
(1-\nu)|y_{k}| + f_{0}$, so if $|y_{k}|>f_{0}/\nu \Rightarrow
|y_{k+1}| < |y_{k}|$. Hence the attractors are found in
$\boldsymbol{\Lambda} \subset \boldsymbol{\Gamma}$, with $
\boldsymbol{\Lambda} = [0, 2\pi] \times [-y_{max},y_{max}]$, where
$y_{max} = f_{0}/\nu$.

\begin{figure}[tb]
\vspace{0.5cm}
\includegraphics[width=.9\columnwidth]{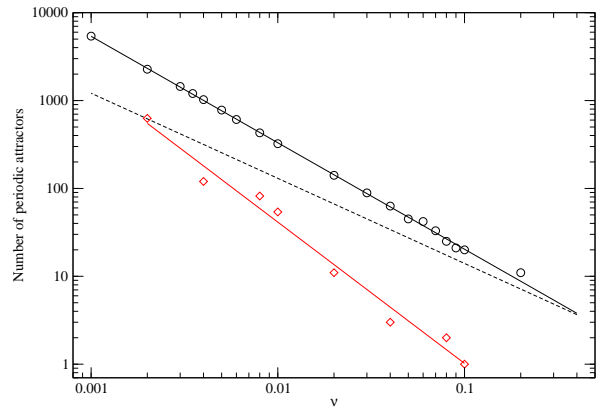}
\caption{(Colour online) Total number of periodic attractors which
  were detected numerically for the single rotor map (black circles),
  and for the H\'{e}non map (red diamonds) for different values of
  $\nu$. The parameters were $f_{0} = 4.0$, for the single rotor map,
  and $A=1.075$ for the H\'{e}non map. It is also shown the power laws fits, the black and the red solid lines for the single
  rotor map and for the H\'{e}non map, respectively; compare to the
  number of period 1 attractors expected from
  Eq. (\ref{eq.urike-attrac}) (black dashed line) for the single rotor
  map. The divergence from this curve (dashed black line) to the other
  (black solid line) is clear, showing the important contribution of
  high period periodic attractors in the regime of small damping for the single
  rotor map.}
\label{fig.compa-attrac}
\end{figure}

In order to detect the attractors, even with very small basins of
attraction, we keep iterating new, randomly chosen initial conditions
$(x_{0},y_{0}) \in \boldsymbol{\Lambda}$ until their trajectories
converge to some periodic attractor. We use up to $2 \times 10^{5}$
iterations; independently of the damping, this provides a convergence of more
than $99.5\%$ of the initial conditions. We have ignored the remaining
trajectories (which may include chaotic orbits), regarding them as not
being statistically representative. It is known that for this and
similar systems, almost all attractors consist of periodic orbits. We
keep iterating new initial conditions until we fail to find new
attractors.  This was determined by the criterion that whenever the
total number of detected attractors did not change for the last
$10^{6}$ initial conditions, we assume that all the detectable
attractors have been found.  This is illustrated in
Fig.~\ref{fig.condxattrac} for $\nu = 0.0035$.  We repeat this
procedure for different values of $\nu$, in order to find how the
total number of periodic attractors changes with damping.  Notice that
the area of the phase space where the trajectories are trapped grows
as we decrease $\nu$; for $\nu = 0.001$ for example, we need $22
\times 10^{6}$ initial conditions to find most attractors in the
system.

The question regarding the finiteness of the number of attractors and
their density in phase space has been considered before and it
is widely regarded as one of the most important open problems still to
be answered in Dynamical Systems Theory~\cite{palis1}. Among many
conjectures, it has been initially proposed that the number of
attractors was infinite~\cite{newhouse74}. Afterwards it has been
proved that this should only hold for a zero measure set of parameter
values, though dense in some interval~\cite{lalli86}. Recently it has
also been conjectured by Palis that the total number of attractors
should be finite~\cite{palis2}.

To investigate this issue using our system (\ref{eq.single-rotor-map})
as a testing ground, we calculate the number of attractors as
described above, and see how this depends on the dissipation $\nu$ as
it is decreased and approaches the conservative limit, i.e. $\nu
\rightarrow \varepsilon >0$. Figure \ref{fig.compa-attrac} shows a
\textit{log-log} plot of our data, yielding a power law describing the
dependence of the total number of periodic attractors on $\nu$.
\noindent Numerical fitting gives us the following law:
\begin{equation}
\label{eq.power-law}
N_{TP_{R}} \approx 1.26 \cdot \nu^{-1.21},
\end{equation}
where $N_{TP_{R}}$ is the total number of detected periodic attractors
for the map~(\ref{eq.single-rotor-map}).

Now we want to compare our power law for the growth of the total number of
periodic attractors with the formula (\ref{eq.urike-attrac})~\cite{feudel96} 
for the number of period 1 attractors. We
observe in Fig.~\ref{fig.compa-attrac} that when the damping is
decreased the contribution of higher period periodic attractors
becomes important for the total number of detected attractors. The
main reason is that the \textit{lifetime} of a stable,
periodic orbit in parameter space increases as damping is reduced. By
\emph{lifetime} of an orbit we mean the range in parameter space for
which the orbit exists, without undergoing any bifurcation. Since the
higher the period of an attractor, the smaller its lifetime interval
in parameter space~\cite{feudel96}, the extension of lifetime for
small damping promotes the overlap of different periodic orbits in the
space of parameters, which was not possible before.
Equation (\ref{eq.urike-attrac}) is only valid within the parameter
intervals where none of the period $1$ attractors has undergone period
doubling~\cite{feudel96}. Since some of the period $1$ attractors may
have gone through such process for the parameters in
Fig.~\ref{fig.compa-attrac}, the difference between the expected
number of period $1$ attractors and the total number of attractors can
be even larger than that inferred from the above argument.

\subsection{The H\'{e}non Map}

We further investigated the growth of the number of periodic
attractors for a map of a different class than the previous one. We
chose the H\'{e}non map in the form~\cite{feudel97}:
\begin{equation}
\label{eq.henon-map}
\begin{split}
x_{n+1} &= A - x^{2}_{n} - (1 - \nu)y_{n}\\
y_{n+1} &= x_{n},
\end{split}
\end{equation}
where $A$ represents the bifurcation or non linearity control
parameter. The parameter $\nu \in [0,1]$ represents the damping
parameter. When $\nu = 1$, the two equations in Eq.~\ref{eq.henon-map}
are no longer coupled, and we obtain the quadratic map. For the other
limit, $\nu=0$, we have a conservative map, hence the determinant of
its Jacobian matrix is equal to $1$. Contrary to the case for the
map~(\ref{eq.single-rotor-map}), the dynamics for the H\'{e}non
map~(\ref{eq.henon-map}) is not \textit{a priori} contained within a
region of the phase space. For a range of parameters, the initial
conditions can either be trapped by some of the coexistent periodic
attractors, or be scattered. In fact, most of the initial conditions
diverge to infinity.

We repeated the procedure applied to the previous map in order
to obtain the number of periodic attractors for the H\'{e}non map. We
fixed $A=1.075$ for the map (\ref{eq.henon-map}), and we iterated a set
of initial conditions in $[-5, 5] \times [-5, 5]$ verifying whether they converged to some periodic
attractor or diverged. Because the number of initial conditions
converging to periodic attractors is much smaller than that diverging,
in order to obtain about $20 \times 10^{6}$ initial conditions
converging to periodic motion, as many as $20 \times 10^{7}$  initial conditions were 
necessary depending on the damping.

Figure~\ref{fig.compa-attrac} shows a
\textit{log-log} plot of our data, yielding a power law describing the
dependence of the total number of periodic attractors on $\nu$.  For
the H\'{e}non map, the numerical fitting produces:
\begin{equation}
\label{eq.power-law-H}
N_{TP_{H}} \approx 0.025 \cdot \nu^{-1.61},
\end{equation}
where $N_{TP_{H}}$ is the total number of detected periodic attractors
for the map~(\ref{eq.henon-map}).

As previously observed~\cite{feudel97}, the H\'{e}non map belongs to a
class of dynamical systems whose the conservative element has few
primary island, which are surrounded by secondary ones. Therefore,
although it also observed the coexistence of periodic attractors for a
range of parameter, it is expected this number to be smaller than that
for the single rotor's family.

\subsection{On the General Behaviour}

Even though we cannot claim the above result to be a mathematical proof,
our results seem to uphold Palis' conjectures~\cite{palis1, palis2},
i.e., for any arbitrarily small $\varepsilon >0$, the limit of
Eqs. (\ref{eq.power-law}), [(\ref{eq.power-law-H})], as $\lim_{\nu
  \rightarrow \varepsilon^{+}}N_{TP_{R,[H]}}$ predicts a finite number
of attractors for the system, though this number can be very high for
small $\varepsilon$ ($N_{TP_{R}}$, and $N_{TP_{H}}$ for the single rotor map, and for the H\'{e}non map, respectively).

Although some attractors may have their basins of attraction so tiny
that even by iterating a huge number of initial conditions is not
enough to numerically detect them. Figure~\ref{fig.condxattrac} shows
that the number of attractors which was detected is asymptotic to a
constant (we obtain a similar figure for the H\'{e}non map). Therefore,
our results indicate that the number of attractors increases as a
power law as the dissipation decreases, being finite for any given
$\nu>0$, though arbitrarily large for $\nu$ approaching zero. Even
though the specific value of the exponent in Eq. (\ref{eq.power-law})
and Eq. (\ref{eq.power-law-H}) may be only valid for the map
(\ref{eq.single-rotor-map}), and the map (\ref{eq.henon-map}),
respectively, the growth of the number of periodic attractors is a very
generic property of weakly dissipative systems~\cite{feudel97} which
should not depend on the details of the system, since in the
low-dissipative limit the system's phase-space approaches that of
Hamiltonian systems, which have universal properties. 
We therefore expect that the same behaviour
to hold for other systems.

The natural question which arises is the following: what happens
to the topology of the phase space as new attractors continually
appear with decreasing damping?


\section{Topology of the phase space}\label{sec.Topology}


We start this section by presenting a qualitative picture of the
transformations occurring in phase space structure when the damping is
decreased. For simplicity, we restrict ourselves to the
map~(\ref{eq.single-rotor-map}). Although one cannot draw definitive
conclusions from this qualitative analysis, it motivates us to some
further quantitative investigation, which will be tackled at the end
of this section and in the next one. In order to get some clue
about which topological changes occur when the damping is decreased,
we begin by looking at the phase space for the system with constant
forcing and high damping. In this case the system has only few
periodic attractors.  We decrease then $\nu$, find the periodic
attractors for the new parameter and plot their basins of
attraction. Since we are now only interested in the general picture,
the number of initial conditions is fixed at $2 \times 10^{6}$.
\begin{widetext}
\begin{figure*}[t]
{\includegraphics[width=.9\linewidth]{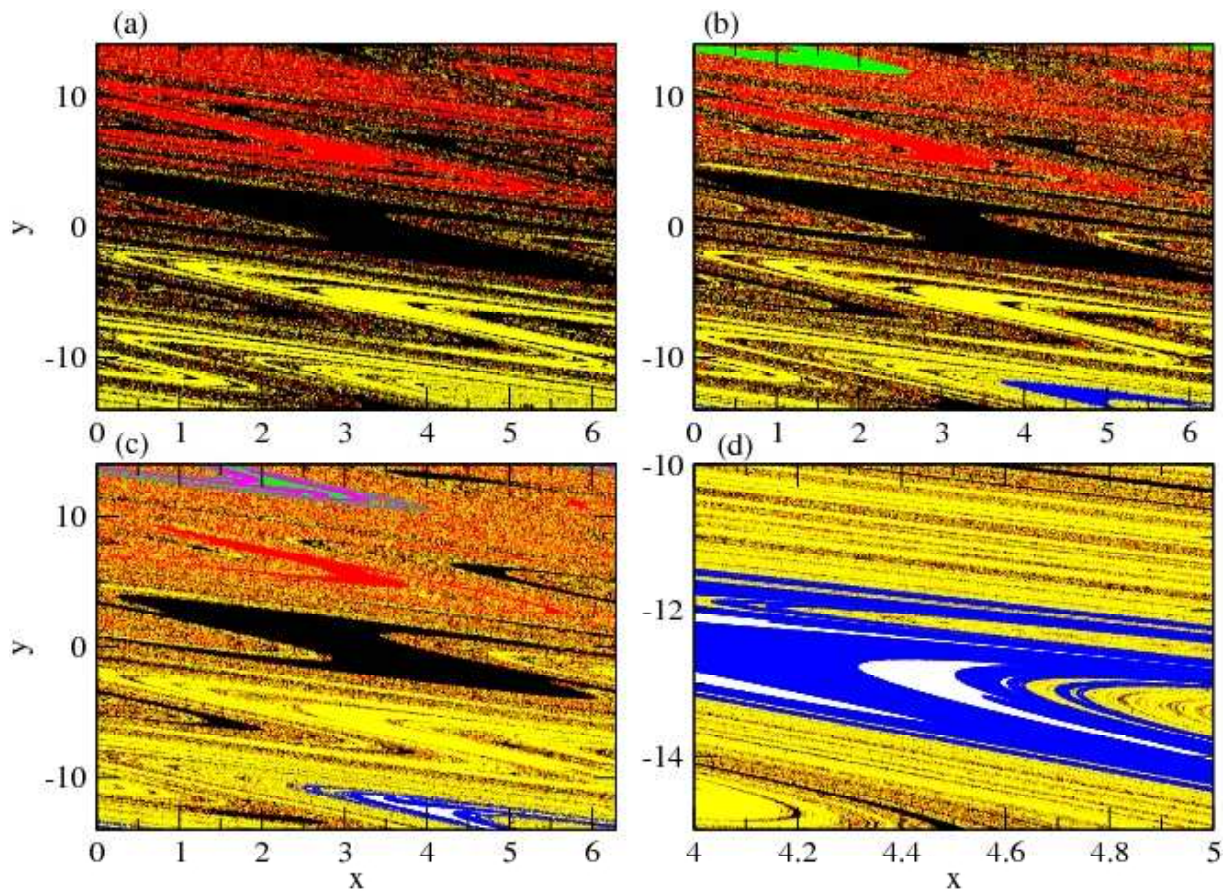}}
\caption{(Colour online) Basins of attraction for $f_{0}=4.0$ at
different values of $\nu$. Each colour represents a set of initial
conditions which converge to one of the attractors. In (a) $\nu=0.32$,
in (b) $\nu=0.3$, in (c) $\nu=0.2415$, and (d) is the blow up of a
region $x \in [4,5], y \in [-15,-10]$ for $\nu=0.2415$. We used $2
\times 10^{6}$ initial conditions for each one of the figures.}
\label{fig.Phasespace}
\end{figure*}
\end{widetext}

\begin{figure}[b]
\includegraphics[width=.9\columnwidth]{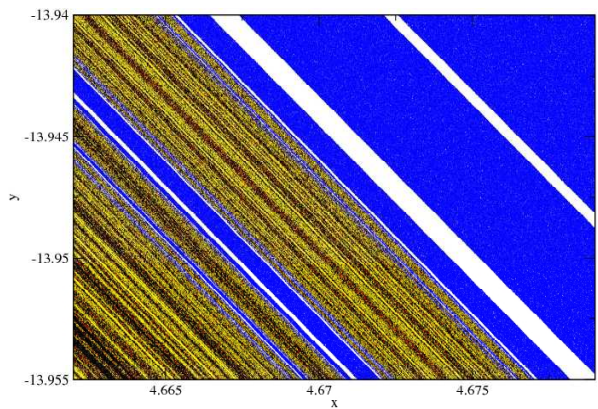}
\caption{(Colour online) Blow up of the region $x \in [4.662, 4.679], y
\in [-13.955,-13.94]$ for $\nu=0.2415$ in Fig.~\ref{fig.Phasespace}
(d), showing the fractal Cantor structure.}
\label{fig.Phaseblowup}
\vspace{0.5cm}
\end{figure}

If we follow the distributions of filaments of the largest basin of
attraction in the phase space, we notice that their structure becomes
more heterogeneous for different regions when the damping is
decreased. This is illustrated in Fig.~\ref{fig.Phasespace}, where
each colour (colour online) represents a set of initial conditions
which converge to one of the attractors, i.e., different colours
represent different basins of attraction.  In
Fig.~\ref{fig.Phasespace} (a), for which $\nu = 0.32$, we notice that
the largest basin of attraction (in black) is spread out over the
phase space in an intricate structure. Moreover, if we compare
distinct regions of the phase space, there is a qualitative difference
in the ``density'' of filaments of the largest basin of attraction. In
particular, the farther the region is from the main attractor, the
thinner the filaments become. In Fig.~\ref{fig.Phasespace} (b), for
$\nu=0.3$, we detect five attractors and the topology of the phase
space, i.e., the spatial distribution of invariant sets, seems to
become more complex. Going even further, for $\nu=0.2415$ in
Fig.~\ref{fig.Phasespace} (c), we have seven attractors in the same
region and the difference of density of those filaments is
clearer. Figure~\ref{fig.Phasespace} (d) is the blow up of the region
$x \in [4,5], y \in [-15,-10]$ for $\nu=0.2415$. The rescaling of the
phase space shows the structure of the basins which seems to form a
self similar pattern, as illustrated in Fig.~\ref{fig.Phaseblowup}.
\begin{figure}[tb]
\includegraphics[width=.9\columnwidth]{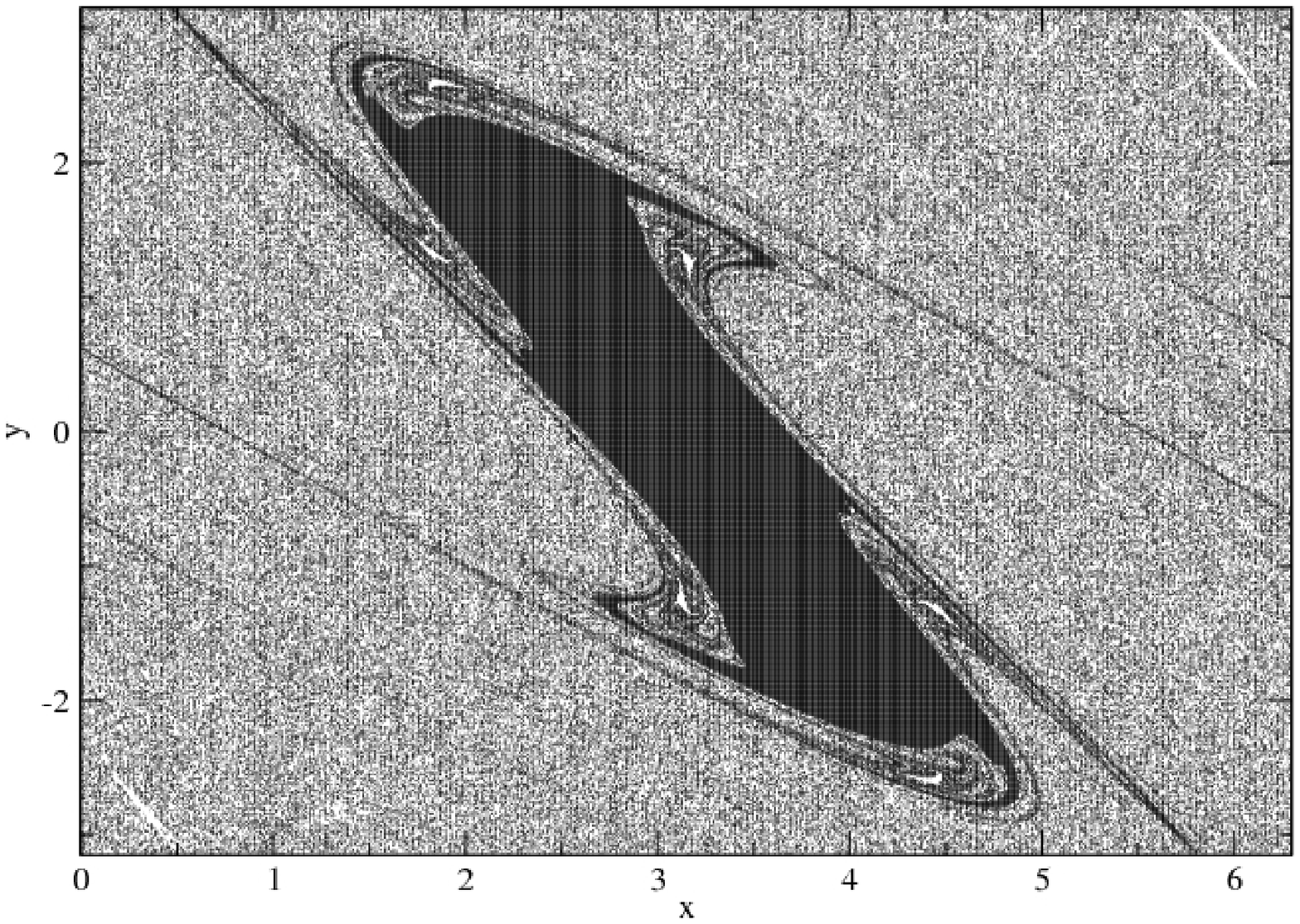}
\caption{The largest basin of attraction for $\nu=0.02$ and $f_{0} =
  4.0$. The black points represent initial conditions which converge
  to the attractor formed from the main island. The initial conditions
  were chosen such that, $(x_{0},y_{0}) \in [0,2\pi] \times [-\pi,
    \pi]$. We used the same parameters as those used in
  Ref~\cite{feudel96}.}
\vspace{0.5cm}
\label{fig.basin}
\end{figure}

As we decrease the damping even further, this process seems to carry
on and the heterogeneity of the phase space keeps
increasing. Furthermore, the difference in density of filaments of
distinct regions of the phase space is seen by looking, for
instance, at the largest basin of attraction for $\nu=0.02$, and
$f_{0} = 4.0$. For these parameters we expect from
Eq. (\ref{eq.power-law}) to find about 140 periodic attractors. The
largest basin of attraction is responsible for about $36\%$ of the
initial conditions $(x_{0},y_{0}) \in [0,2\pi] \times [-\pi,
\pi]$~\cite{feudel96}. This is illustrated in Fig.~\ref{fig.basin},
which has been done by the iteration of an ensemble of $10^{6}$
initial conditions.

The structures we just presented are nothing but manifestations of some of the
underlying invariant sets. That is, the dynamics has multiple periodic attractors, and the closure of the set of initial conditions that approaches a give attractor is its basin of attraction. The boundary which separates different basins of attraction is the basin boundary. This basin boundary is the stable
manifold of an invariant set.  If we define the boundary
$B$ in some region of the phase space embedded in the boundary as
$S\cap B$, where $S$ denotes some open set in phase space which
contains part of the boundary $B$, we expect for a given $\nu$ the fractal dimension of the basin boundary, $dim(S \cap B)$,
to be constant at different regions of the phase space,
$\boldsymbol{R} \subset \boldsymbol{\Gamma}$, as it has been
conjectured in Ref.~\cite{grebogi88}. Nevertheless, for physically
relevant scales, the heterogeneity in the distribution of the basins
over the phase space, and hence the spatially heterogeneous
distribution of invariant manifolds, suggests that course-grained
measurements of quantities such as the fractal dimension should lead to
different results depending on the region of the phase space and on
the scale under consideration. The reason is that realistic
measurements never take the limit of infinitely small scales, but must
have a finite lower scale.  Previous works in the context of
Hamiltonian systems show that when the phase space has a heterogeneous
structure such as in this case, one needs to go to exceedingly small
scales to approach the mathematical value of the fractal dimension,
which is unique and independent of the portion of the phase space used
to calculate it.  For physically realistic scales, an approximation to
the fractal dimension (and similar quantities) is more appropriate to
describe the system's dynamics; this approximation, called the
\emph{effective fractal dimension}, depends on the position as
well as on the scale under consideration. This effective fractal dimension will be presented in more detail in the
next section.

A finite-scale approximation of the fractal dimension of the basin
boundary can be estimated using the uncertainty
method~\cite{grebogi83}. It consists in iterating an ensemble of
initial conditions $(x_{0},y_{0})$ and checking to which attractor they
converge. Then we add a small perturbation $\varepsilon$ to every
initial condition, say $(x_{0},y_{0}+\varepsilon)$, and check whether they
change from one basin to another. We count the fraction of initial
conditions that have changed their asymptotic state after being
perturbed in this manner. This fraction, $f(\varepsilon)$ of
$\varepsilon$-uncertain points is the fraction of initial
conditions that change basins under an $\varepsilon$-size
perturbation, and it scales as $f(\varepsilon) \sim \varepsilon^{\alpha}$,
where $\alpha$ is related to the box counting dimension $d$ of the
basin boundary by $\alpha = D - d$, where D is the dimension of the
phase space (For the map (\ref{eq.single-rotor-map}) we have $D=2$).

\begin{figure}[tb]
\vspace{0.5cm}
\includegraphics[width=\columnwidth]{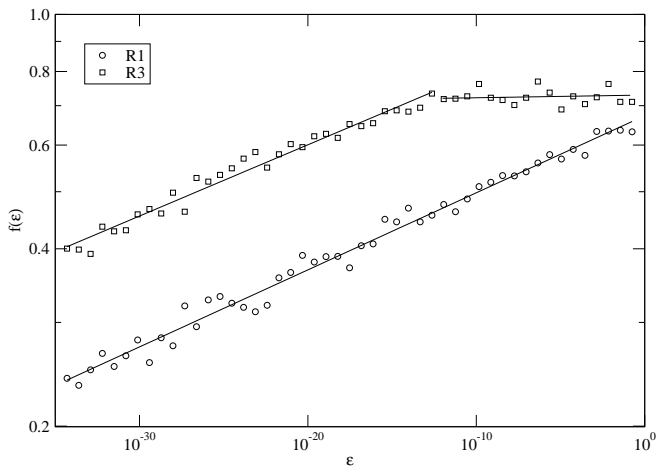}
\caption{Fraction of uncertain initial conditions $f(\varepsilon)$ as
  a function of the perturbation size $\varepsilon$ for two different
  regions of the phase space. The regions, $\boldsymbol{R}_{i} \subset
  \boldsymbol{\Gamma}$, were defined as $\boldsymbol{R}_{1} \equiv
  (x,y) \in [0,1.5] \times [-1.0, -2.5]$, and $\boldsymbol{R}_{3}
  \equiv (x,y)\in [1.0,2.5] \times [198.5, 200.0]$. The utilised
  parameters were $f_{0}=4.0$, and $\nu = 0.08$.}
\label{fig.dimension}
\end{figure}

In Fig.~\ref{fig.dimension} we show the fraction of ``uncertain''
initial conditions as function of the size of perturbation
$\varepsilon$~\cite{computation}. The slope of the curves are related
to the fractal dimension of the basin boundary.

\noindent The exponents $\alpha = \Delta\ln f(\varepsilon) / \Delta
\ln \varepsilon$ can be computed over some decades. However, we notice
in Fig.~\ref{fig.dimension} that the slope assumes different values
for distinct regions of the phase space. Such regions,
$\boldsymbol{R}_{i} \subset \boldsymbol{\Gamma}$, are defined as
$\boldsymbol{R}_{1} \equiv (x,y) \in [0,1.5] \times [-1.0, -2.5]$, and
$\boldsymbol{R}_{3} \equiv (x,y)\in [1.0,2.5] \times [198.5, 200.0]$,
and they were chosen with the same Lebesgue measure. We chose three
different regions of the phase space, located at different distances
from the main attractor.  We remark that, although such regions may be
out of $\boldsymbol{\Lambda} \subset \boldsymbol{\Gamma}$ defined in
section~\ref{sec.N-of-Att}, it does not follow that they do not
contain boundaries. This is because regions $\boldsymbol{\Lambda}
\subset \boldsymbol{\Gamma}$, defined for each value of $\nu$, are
portions of the phase space where the attractors are to be found;
their basins, on the other hand, are expected to be extended throughout
the phase space.  Since the density of periodic attractors
in phase space decreases when we look at regions farther away
from the main attractor, we expect the stable manifold to be denser
around the central part of the phase space. Furthermore, given that we
expect the dissipative regime to approach the complex structures of
the Hamiltonian dynamics at the zero dissipation limit, it is sensible
to expect a distribution of manifolds which are similar for
conservative and for low dissipative dynamics. For Hamiltonian
dynamics, the density of manifolds decreases as one looks at regions
farther from the main island~\cite{motter05}.

Looking at Fig.~\ref{fig.dimension}, it is clear that for $\varepsilon > 10^{-13}$, the estimated
fractal dimension is larger for the region $\boldsymbol{R}_{3}$, where
we expect the stable manifold to be less dense. Only when we go down to
$\varepsilon < 10^{-13}$ does the slope for the region
$\boldsymbol{R}_{3}$ converges to the same value as the one of the region
$\boldsymbol{R}_{1}$: for small enough $\varepsilon$, the
slope eventually converges to the true fractal dimension, which has a
unique value~\cite{grebogi88}. We have computed $f(\varepsilon)$ for
even smaller values of $\nu$, and we have observed the same
behaviour. However, for values as small as $\nu=0.02$, even going down
to $\varepsilon = 10^{-35}$ is not enough to obtain a
convergence to a unique value of the estimated dimension in different
regions. This shows that an effective dimension is indeed the relevant
physical quantity to be considered also for weakly dissipative
systems.


\section{Effective Fractal Dimension}\label{sec.Effective}


From the theoretical results regarding the fractal dimension of the basin
boundary~\cite{grebogi88}, we expect to have the same
fractal dimension for different regions of the phase space. However,
the above discussion and results suggest that, for realistic scales,
the behaviour for weakly dissipative systems can be quite different
from the asymptotic one. Therefore it is sensible to describe the
system in terms of \emph{effective} invariants, a concept which has
been used mostly in the context of non-hyperbolic Hamiltonian dynamical
systems.~\cite{motter05}. As an example of effective invariants we use
the definition of effective fractal dimension for a $D$-dimensional
system, which depends on the region $\boldsymbol{R} \subset
\boldsymbol{\Gamma}$, and on the scale under
consideration~\cite{comment}. It is defined as
\begin{equation}
D_{eff}(R,\varepsilon) = \left.D - \frac{d \ln f(\varepsilon')}{d \ln
\varepsilon '} \right|_{\varepsilon ' = \varepsilon} \quad,
\end{equation} 
where $f(\varepsilon')$ is the total uncertain phase space volume
estimated as $\varepsilon'^{D} N(\varepsilon')$, and $N(\varepsilon')$
is the number of hypercubes of size $\varepsilon'$ necessary to cover
$S \cap B$.

We have computed the effective fractal dimension of the basin boundary
for finite scales, for different values of $\nu$, and for distinct
regions in the phase space. The regions, $\boldsymbol{R}_{i} \subset
\boldsymbol{\Gamma}$, were chosen to be $\boldsymbol{R}_{1} \equiv (x,y)
\in [0,1.5] \times [-1.0, -2.5]$, $\boldsymbol{R}_{2} \equiv (x,y)\in
[1.0,2.5] \times [18.5, 20.0]$, and $\boldsymbol{R}_{3} \equiv
(x,y)\in [1.0,2.5] \times [198.5, 200.0]$.
\begin{figure}[tb]
\includegraphics[width=\columnwidth]{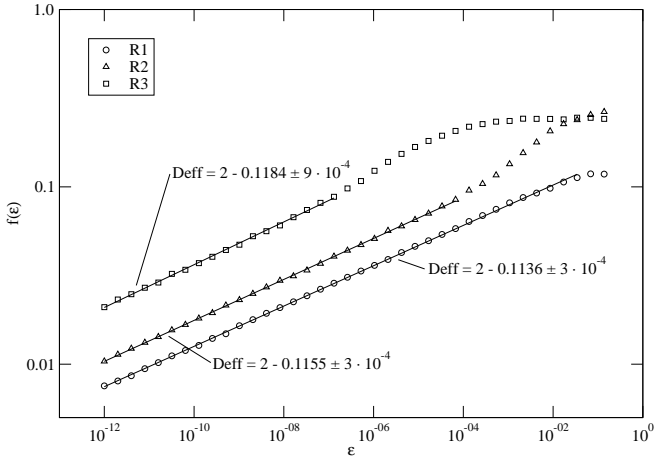}
\caption{Fraction of uncertain initial conditions $f(\varepsilon)$
scaling with the perturbation $\varepsilon$ for three different
regions of the phase space. The regions, $\boldsymbol{R}_{i} \subset
\boldsymbol{\Gamma}$, were defined as $\boldsymbol{R}_{1} \equiv (x,y)
\in [0,1.5] \times [-1.0, -2.5]$, $\boldsymbol{R}_{2} \equiv (x,y)\in
[1.0,2.5] \times [18.5, 20.0]$, and $\boldsymbol{R}_{3} \equiv
(x,y)\in [1.0,2.5] \times [198.5, 200.0]$. The utilised parameters
were $f_{0}=4.0$, and $\nu = 0.3$.}
\label{fig.dimension3}
\vspace{0.5cm}
\end{figure}
\begin{figure}[tb]
\includegraphics[width=\columnwidth]{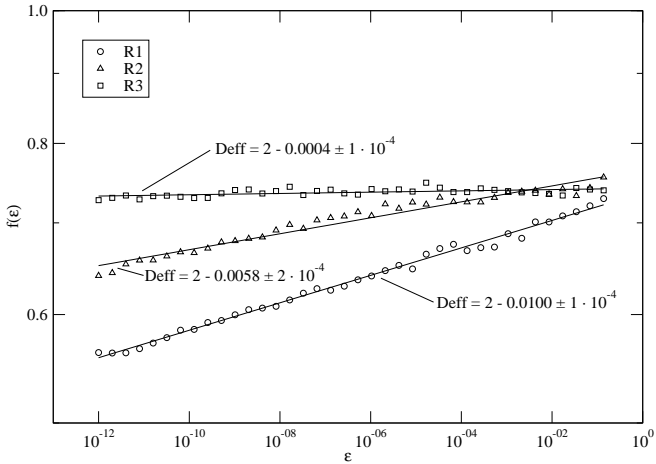}
\caption{Fraction of uncertain initial conditions $f(\varepsilon)$ scaling with
the perturbation $\varepsilon$ for three different regions of the
phase space The regions are the same as in
Fig.\ref{fig.dimension3}. The utilised parameters were $f_{0}=4.0$,
and $\nu = 0.07$.}
\label{fig.dimension07}
\vspace{0.5cm}
\end{figure}
\begin{figure}[tb]
\includegraphics[width=\columnwidth]{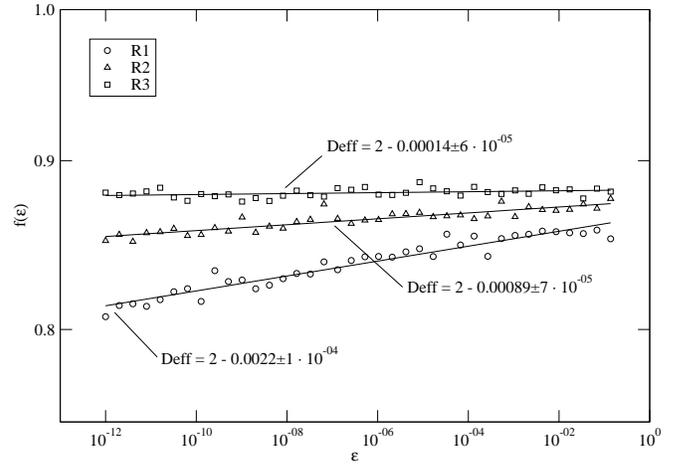}
\caption{Fraction of uncertain initial conditions $f(\varepsilon)$ scaling with
the perturbation $\varepsilon$ for three different regions of the
phase space. The regions are the same as in
Fig.\ref{fig.dimension3}. The utilised parameters were $f_{0}=4.0$,
and $\nu = 0.02$.}
\label{fig.dimension02}
\end{figure}

We have extracted the effective fractal dimension from
Fig.~\ref{fig.dimension3} for $\nu = 0.3$ and $f_{0}=4.0$, where the
system has apparently $5$ periodic attractors. We notice that for
$\boldsymbol{R}_{3}$, where the stable manifold is less dense,
starting from $\varepsilon = 10^{-1}$ , the value of the effective
fractal dimension is larger, but after few decades it converges to
nearly the same as for $\boldsymbol{R}_{1}$, i.e., $D_{eff} = 1.89$,
where the stable manifold is denser. The same phenomenon is observed
for $\boldsymbol{R}_{2}$, though the convergence is faster than for
$\boldsymbol{R}_{3}$. When we decrease $\nu$, hence increasing the
number of attractors, even for $\varepsilon = 10^{-12}$, which is
almost in the limit of normal computation and far beyond realistic
measurement capacity, we do observe different values of effective
fractal dimension for different regions, as it is shown in
Fig.~\ref{fig.dimension07} and Fig.~\ref{fig.dimension02}, for $\nu =
0.07$ and $\nu = 0.02$, respectively. We also notice that for a given
region the exponent $\alpha$ becomes smaller as the damping is
decreased. This indicates that for such weakly dissipative systems,
the fractal dimension is very close to the dimension of the phase
space. This is, of course, in accordance with the fact that for
Hamiltonian systems ($\nu=0$), the fractal dimension assumes its
maximum value, i.e., $D$.
%
%
%

 

\section{Conclusion}\label{Sec.conclu}

We have shown that the dynamics of weakly dissipative dynamical
systems can be quite different from either the strongly dissipative
systems or the Hamiltonian ones. When the damping is decreased, the number
of periodic attractors increases. We have shown that the growth of the
total number of attractors is described by a power law. In general,
the interval in the space of parameters in which an attractor exists
is smaller for high period periodic attractors.  For small damping,
attractors with different periods coexist in phase space, and the
contribution of high period periodic attractors to the dynamics becomes
important.  Although the total number of attractors in the system is
very high for low damping, our results strongly suggests that it
remains finite for $\nu>0$, supporting Palis' conjecture. The
formation of new attractors as the damping decreases changes the
topology of the phase space considerably.  For low damping, the
dynamics is better characterised by effective dynamical invariants, in
a similar way to the case of non-hyperbolic Hamiltonian systems. These
effective dynamical invariant sets depend on the scale and they differ from one
region to the other in the phase space. In particular, the effective
fractal dimension of the basin boundary is larger for regions where
the stable manifold is less dense. For a given region, the smaller the
damping, the closer the effective fractal dimension is to the
dimension of the phase space. Although we have used a specific example
to illustrate our ideas, our results are generic since we have deal
only with general characteristics of dynamical systems close to
Hamiltonian case~\cite{palis2}.

\begin{acknowledgments} 

The authors thank E. G. Altmann for the careful reading of the
manuscript and useful suggestions. C. S. R. thanks I. J. Rodrigues
for illuminating discussion, and takes the opportunity to pay his
tribute to J. S. Rodrigues (\textit{in memorian}), who gave him the
first scientific hints. This work was supported by the School of
Natural Sciences, University of Aberdeen.

\end{acknowledgments}


\section*{Appendix}
\subsubsection*{On Periodic Attractors}

Some of the most important characteristics of the dynamics are related
to invariant subsets in the phase space, which attract their
neighbouring points. These sets are called \textit{attracting sets},
or \textit{attractors}, depending on the context. There are various
definitions for \textit{attractor}, the main difference is related
to which points in the neighbourhood must approach the set. We use the
definition given for a \textit{diffeomorphism} $f: \mathbb{R}^{N}
\rightarrow \mathbb{R}^{N}$, that is $f$ and its inverse are
differentiable and the partial derivatives are continuous. The main
source of our definition is Ref.~\cite{ruelle}.

The dynamics takes place in a phase space $\boldsymbol{\Gamma}
\subset \mathbb{R}^{N}$, where there is the notion of distance
$dist(\textbf{x}_{1},\textbf{x}_{2})$ between points $\textbf{x}_{1}$
and $\textbf{x}_{2}$, both belonging to $\boldsymbol{\Gamma}$. For
example, $dist(\textbf{x}_{1},\textbf{x}_{2})$ can be the
\textit{Euclidean distance}~\cite{note}. We say that a bounded and
limited region of the phase space, i.e. $U_{\boldsymbol{A}}
\subset \boldsymbol{\Gamma}$ is called a \textit{trapping region} for
$f$ if $f(U_{\boldsymbol{A}}) \subset int(U_{\boldsymbol{A}})$,
where $int(U_{\boldsymbol{A}})$ stands for the interior of
$U_{\boldsymbol{A}}$. Another important concept for the definition of
\textit{attractors} is the notion of $\epsilon$-\textit{chain}. An
$\epsilon$-\textit{chain} of length $n$ from $\textbf{x}$ to
$\textbf{y}$, for a map $f$ is a sequence $\{ \textbf{x} =
\textbf{x}_{0}, \ldots, \textbf{x}_{n} = \textbf{y}\}$, such that for
all $1 \leq j \leq n$, we have $dist(f(\textbf{x}_{j-1}),
\textbf{x}_{j}) < \epsilon$. Now we are ready to state a formal
definition of \textit{periodic attractors}. If forward iterating our
map $f$ in a trapping region, i.e. if we take the set $\boldsymbol{A} =
\bigcap_{n \geq 0} f^{n}(U_{\boldsymbol{A}})$, and we
eventually end up in an unique closed sequence of points such as these
points are an $\epsilon$-\textit{chain} no matter how small
$\epsilon>0$ is, we have precisely what is defined as
\textit{attractor}. In some other words, a periodic attractor is the
unique sequence of points $\{ \textbf{x}_{1}, \textbf{x}_{2}, \ldots,
\textbf{x}_{p}, \textbf{x}_{1}\}$ remaining by iterating a trapping
region, such that $dist(f(\textbf{x}_{j-1}), \textbf{x}_{j}) <
\epsilon$, for all $\epsilon>0$. In this case, we have \textit{period}
$p$. Notice that the attractivity of the attractor is ensured by
requiring the existence of the trapping region. Furthermore, we remark
that, finding multiple periodic attractors require the coexistence of
different trapping regions.

\begin{figure}[tb]
\vspace{0.1cm}
\includegraphics[width=.7\columnwidth]{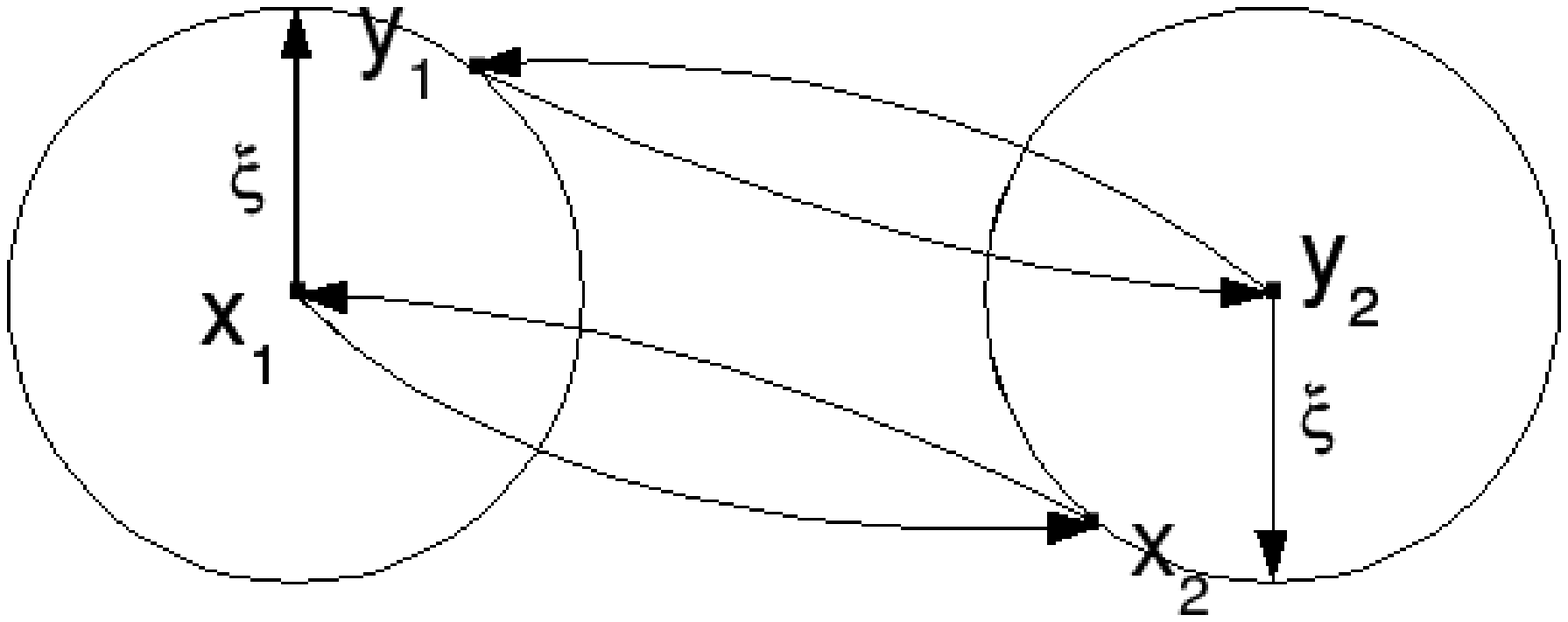}
\caption{Illustration of the dependence of the number of periodic
  attractors with the precision. The figure shows attractor $x_{i}$,
  and $y_{i}$, with $i=1$, or $2$. Hence, the attractors are of
  period two. If on the detection of such attractors, $\epsilon > \xi$
  is used, only one attractor of period $2$ is detected. On the other
  hand, for $\epsilon < \xi$, we are able to distinguish them.}
\label{fig.illust}
\end{figure}

Although the formal definition of attractor requires the existence of such sequence
of points, the $\epsilon$-\textit{chain}, for all $\epsilon>0$, and infinitely many iterations, both experimentally, and numerically, we are
constrained by limited precision. Hence we are forced to use a mild
condition on $\epsilon$. Therefore, the number of periodic attractors
obtained in realistic investigation intrinsically depends on the
precision that we choose. We give a picturesque illustration of it in
Fig~\ref{fig.illust}. On determining of the number of periodic
attractors in this paper, we have used $\epsilon =
10^{-10}$. Nevertheless, it does not minimise the importance of ours
findings, showing fundamental characteristics of the dynamics within
realistic scale.

We also use a more general definition of \textit{basin of
  attraction}. The \textit{basin of attraction} of some trapping
region $U$ is the set of positive Lebesgue measure of
initial conditions, whose orbits eventually enter
$U$, as defined in Ref.~\cite{nusse97}. Note that this
slightly different definition avoids problems of coherence for not
having an infinitely small $\epsilon$. Because any periodic
attractors is contained in some trapping region, hereafter we often
mention the \textit{basin of attraction} of some attractor, and it is
understood the basin of a given trapping region containing the attractor under consideration.



\begin{thebibliography}{10}

\bibitem{neural} N. Nagao, H. Nishimura, and N. Matsui,
\newblock {\em Neural Processing Lett} \textbf{12}, 267 (2000).

\bibitem{neural2}
S. J. Schiff and K. Jerger and D. H. Duong and et al.,
\newblock {\em Nature } \textbf{370}, 615 (1994).

\bibitem{kimPRE02}
O. Peters, and K. Christensen,
\newblock {\em Phys. Rev. E} \textbf{66}, 036120 (2002).

\bibitem{kimPRL02}
P. Bak, K. Christensen, L. Danon, and T. Scanlon,
\newblock {\em Phys. Rev. Lett} \textbf{88}, 178501-1 (2002).

\bibitem{chaos04}
M. Anghel,
\newblock {\em Chaos Solit \& Frac} \textbf{19}, 399 (2004).

\bibitem{wang97}
W. F. Wang, and P. P. Ong,
\newblock {\em Phys. Rev. A} \textbf{55}, 1522 (1997).

\bibitem{motter05}
A. Motter, A.~P.~S. de Moura, C. Grebogi, and H. Kantz,
\newblock {\em Phys. Rev. E} \textbf{65}, 026120 (2005).

\bibitem{meiss83}
J. D. Meiss, J. R. Cary, J. D. Crawford, C. Grebogi, A. K. Kaufman,
and H. D. I. Abarbanel, Physica D \textbf{6}, 375 (1983).

\bibitem{altmann06}
E. G. Altmann, A. Motter, and H. Kantz,
\newblock {\em Phys. Rev. E} \textbf{73}, 026207 (2006).

\bibitem{feudel96}
U. Feudel, C. Grebogi, B. R. Hunt, and J. A. Yorke,
\newblock {\em Phys. Rev. E} \textbf{54}, 71 (1996).

\bibitem{palis1}
J. Palis,
\newblock {\em A global perspective for non-conservative dynamics},
Institute of Pure and Applied Mathematics (IMPA), preprint, 2005. See
also references therein.

\bibitem{palis2}
J. Palis,
\newblock {\em Ast{\'e}risque} \textbf{261}, 335 (2000).

\bibitem{Zaslavskii1978}
G. M. Zaslavskii,
\newblock {\em Phys. Lett. A} \textbf{69}, 145 (1978).

\bibitem{Chirikov}
B. Chirikov,
\newblock {\em Phys. Rep. A} \textbf{52}, 265 (1979).

\bibitem{newhouse74}
S. E. Newhouse,
\newblock {\em Topology} \textbf{13}, 9 (1974).

\bibitem{lalli86}
L. Tedeschini-Lalli, and J. A. Yorke,
\newblock {\em Commun. Math. Phys} \textbf{106}, 635 (1986).


\bibitem{feudel97}
U. Feudel, and C. Grebogi,
\newblock {\em Chaos} \textbf{7}, 597 (1997);
\textit{ibid}, Phys. Rev. Lett. \textbf{91}, 134102 (2003).

\bibitem{grebogi88}
C. Grebogi, H. E. Nusse, E. Ott, and J. A. Yorke,
\newblock {\em in Lectures Notes in Mathematics} \textbf{1342}, 220
Ed by J. C. Alexander, Springer-Verlag, New York, 1988.  See
also references therein.

\bibitem{grebogi83} C. Grebogi, S. W. McDonald, E. Ott, and
J. A. Yorke, 
\newblock {\em Phys. Lett} \textbf{99A}, 415 (1983).

\bibitem{computation} \newblock {The numerical simulations for
$\varepsilon \leq 10^{-13}$ were carried out with high precision by
using the MPFR C library for arbitrary precision~\cite{mpfr}, which
seemed to be the best choice for trigonometric functions. However, the
computational cost is relatively high in comparison to standard C
precision. For the exponent uncertainty, we have iterated a sufficient
number of initial conditions, in order to reach $400$ uncertain
points. Using standard precision, we have fixed this limit up to
$10000$.}

\bibitem{mpfr} L. Fousse, G. Hanrot, V. Lef\`{e}vre, P. P\'{e}lissier,
and P. Zimmermann, 
\newblock {MPFR: A multiple-precision binary
floating-point library with correct rounding.}
\newblock{\em ACM Trans. Math. Softw.} \textbf{33}, 13 (2007).

\bibitem{comment} \newblock {Effective fractal dimension has been
defined in a similar way for Hamiltonian system~\cite{motter05}.}

\bibitem{ruelle} D. Ruelle, 
\newblock {\em Comm. in Mathematical Physics} \textbf{82}, 137 (1981).

\bibitem{note} \newblock {In fact this definition is general for a
  $C^{r}$-diffeomorphism $f: M \rightarrow M$, where $M$ is some
  smooth manifold.}

\bibitem{nusse97} H. E. Nusse, and J. A. Yorke.  \newblock {\em
  Ergod. Th. \& Dynam. Sys}, 17:463, 1997; H. E. Nusse, and
  J. A. Yorke.  \newblock {\em Physica D}, 90: 242, 1996; H. E. Nusse,
  and J. A. Yorke.  \newblock {\em Phys. Rev. Lett.}, 84: 627, 2000;

\end{thebibliography}
\end{document}